1# PREDICTING AFROBEATS HIT SONGS USING SPOTIFY DATA

**Adewale Adeagbo**
asadeagbo@go.stcloudstate.edu## ABSTRACT

This study approached the Hit Song Science problem with the aim of predicting which songs in the Afrobeats genre will become popular among Spotify listeners. A dataset of 2063 songs was generated through the Spotify Web API, with the provided audio features. Random Forest and Gradient Boosting algorithms proved to be successful with approximately F1 scores of 86%.## I.  INTRODUCTION

Afrobeats is a music genre from West Africa with a style that is rooted in a diverse fusion of sound with influences ranging from hip hop, soca, jazz, R&B, pop, to traditional West African genres like fuji, juju, and highlife. The success of the genre in the last ten years has birthed a global musical movement raging through radio stations in New York, pubs in Paris, to beach parties in the Caribbean.

This study aims to contribute to literature on Hit Song Science (HSS), which is an active research area in the field of Music Information Retrieval (MIR). The result presented here is particularly relevant to record labels, recording artistes, and music industry executives who are looking for an edge in a competitive music industry.

## II.  RELATED WORK

This research is first of its kind within the Afrobeats genre and the idea for an Afrobeats focused HHS research follows from previous studies that have used Spotify and Billboard data to predict Billboard Hot 100 hits. [1][2]

## III.  METHODS

**Dataset and features** A dataset with 2063 Afrobeats songs released from 2010 till date was collected from Spotify's API endpoints.

Since we are concerned with topic of hit songs, the songs included in this dataset were sourced from Spotify playlists created by individual users or Spotify "editors".

Playlists on Spotify is the rail track through which songs are transported from obscurity to mass appeal – a strategy referred to as "playlisting". [3] All of the songs used in this research were already on existing playlists and duplicates were eliminated.

**Features** Spotify runs a suite of audio analysis algorithms on tracks in their catalog to extract some high-level acoustic attributes, some of which are well known musical features. The information offered from the algorithm analysis is the basis of this research.

A target variable called ***Hit*** was created and labeled '1' for a hit song or '0' for a non-hit song. Additional 13 features were selected from data provided in Spotify's *audio-features* and *track* endpoints.

In the absence of an authoritative music industry record chart for Afrobeats like Billboard Hot 100 which was used in previous studies, the main challenge in this research was answering the question: *what qualifies as a hit song?*

The target variable *hit* was created from the Spotify's Track API *popularity* endpoint.

Spotify assigns a value between 0 and 100 to each track based on the popularity of the track. According to Spotify, this value is calculated by an algorithm and *"is based, in the most part, on the total number of plays the track has had and how recent those plays are"*.

This researcher does not objectively consider himself an authoritative figure on music charts. Since the full details of the ranking algorithm is not made available to the public, an exhaustive reading of Spotify's documentation[4] indicates:

i. Songs that are being played a lot now will have a higher value than songs that were played a lot in the past.
ii. Songs released before the popularity API endpoint was released have no popularity ranking.
iii. If they do have, that means the particular 'old song' is an evergreen getting massive replay.
iv. The algorithm keeps it pretty balanced – this researcher's familiarity with Afrobeats, and the insight gained from looking at values assigned to the biggest songs makes this a safe assumption.

See figure 1 for the distribution of the popularity variable in the dataset.

**Figure 1:**

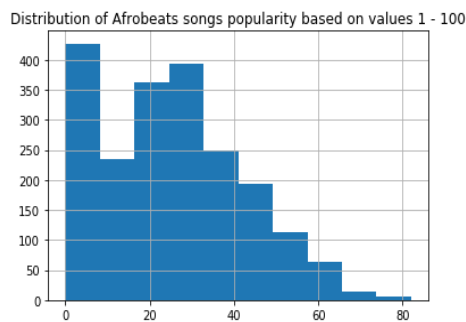

The maximum assigned value is 82 and the minimum is 0, the mean value is 25. Some of the biggest radio hits have values between 50 and 66. Since we're trying to predict hits *(i.e. the cream of the crop)*, we cannot use the mean value as a threshold for what qualifies as a hit song.

Based on the insight gained from an exploratory analysis of the dataset and qualitative factors, we select **47** as the base popularity value for what could be considered a hit song today.

Following this, a song with popularity value greater than 47 is labeled '1' and considered a hit while songs below 47 are labeled '0'. This gives us the values for the target variable ***Hit***.

The description of the 13 features or dependent variables used in this research is given below [5]:

- Danceability: describes how suitable a track is for dancing based on a combination of musical elements including rhythm stability, beat strength, and overall regularity.

- Energy: represents a perceptual measure of intensity and activity. Typically, energetic tracks feel fast, loud, and noisy.

- Key: The estimated overall key of the track. Integers map to pitches using standard Pitch Class notation. E.g. 0 = C, 1 = C♯/D♭, 2 = D, and so on. If no key was detected, the value is -1.

- Loudness: The overall loudness of a track in decibels (dB). Loudness values are averaged across the entire track and are useful for comparing relative loudness of tracks.

- Mode: Mode indicates the modality (major or minor) of a track, the type of scale from which its melodic content is derived. Major is represented by 1 and minor is 0.

- Speechiness: detects the presence of spoken words in a track. The more exclusively speech-like the recording (e.g. talk show, audio book), the closer to 1.0 the attribute value. Values between 0.33 and 0.66 describe tracks that may contain both music and speech, including such cases as rap music.

- Acousticness: A confidence measure from 0.0 to 1.0 of whether the track is acoustic.

- Instrumentalness: Predicts whether a track contains no vocals. "Ooh" and "aah" sounds are treated as instrumental in this context. Rap or spoken word tracks are clearly "vocal".

- Liveness: Detects the presence of an audience in the recording. Higher liveness values represent an increased probability that the track was performed live.

- Valence: A measure describing the musical positiveness conveyed by a track. Tracks with high valence sound more positive, while tracks with low valence sound more negative.

- Tempo: The overall estimated tempo of a track in beats per minute (BPM). In musical terminology, tempo is the speed or pace of a given piece and derives directly from the average beat duration.

- Duration ms: The duration of the track in milliseconds.

- Time signature: An estimated overall time signature of a track. The time signature (meter) is a notational convention to specify how many beats are in each bar (or measure).

**Dataset** The shape of the dataset is [2063,13], which we split into training, test, and validation sets. The split ratio is 80% training – 20% test; with an additional 75% training – 25% validation split on the original training set. [6]

The distribution of the target variable indicates that we have class imbalance in the dataset, with 1826 non hits and 237 hits. See figure 2.

**Figure 2:**

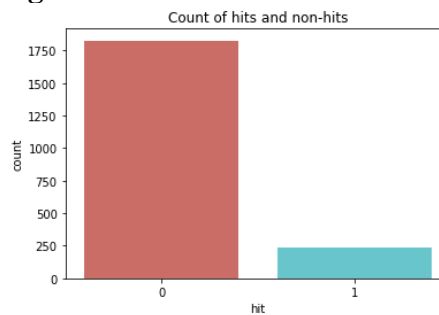

*Figure 2: Distribution of the target variable*

Observed imbalance in the distribution of the target variable means that the result reported in this study must include F1 score of the confusion matrix, along with precision, ROC curve, and recall values.[6]

## IV. ALGORITHMS

To predict whether a given Afrobeats song is going to be a hit or not, since this is essentially a classification task, we use five different models:

i. Logistic Regression (LR)
ii. Decision Tree (Decision Tree)
iii. Random Forest Classifier (RF)
iv. Gradient Boosting - XGBoost (XGB)
v. Neural Network (NN)

**Logistic regression** (LR)
We use a binary logistic regression with *sigmoid* activation function to constrain our probability estimate between 0 and 1. Parameter estimation is done using Maximum Likelihood Estimation (MLE).

The dependent variable under the LR model need not be normally distributed. Feature scaling is applied to ensure the features update at the same speed.

**Decision Tree** (DT)
A decision tree is implemented to investigate how well the hierarchical structure of the trees learns signal from both classes.

**Random Forest** (RF)
This is one of the more popular tree ensemble methods and it often perform better on imbalanced datasets because it aggregates many decision trees to limit overfitting and error due to bias. This model is implemented with *100 estimators*.

**Gradient Boosting** (XGB)
We use the gradient boosting technique package developed by Tianqi Chen and Carlos Guestrin[7]. XGB is implemented with *100 estimators* as well.

**Neural Network** (NN)
We implement a neural network with the Keras Classifier wrapper and a grid search parameter tuning approach because of the modest datasize.

The architecture includes an outer layer with ReLU activation function, one hidden layer with ReLU activation function, and an output layer with sigmoid activation function. We use the Binary Cross-Entropy Loss function and Adam optimizer.

## V. RESULTS

In reporting the performance of each of our models on the training and validation sets, we look beyond the given accuracy score and give equal importance to precision, recall, and also the F1 score.[8]

The metric we are trying to optimize in this study is the precision score because a false positive prediction could prove costly for a record label if they commit resources to a non-hit song that is falsely classified as a hit.

**Table 1**: Model Results for test set

| Model | Accuracy | Precision | Recall | F1 score |
|---|---|---|---|---|
| LR | 0.87 | 0.80 | 0.87 | 0.82 |
| DT | 0.82 | 0.81 | 0.82 | 0.82 |
| RF | 0.88 | 0.85 | 0.88 | 0.83 |
| XGB | 0.88 | 0.83 | 0.88 | 0.83 |

**Table 2**: Model Results for validation set

| Model | Accuracy | Precision | Recall | F1 score |
|---|---|---|---|---|
| LR | 0.90 | 0.84 | 0.90 | 0.85 |
| DT | 0.82 | 0.83 | 0.82 | 0.83 |
| RF | 0.90 | 0.86 | 0.90 | 0.86 |
| XGB | 0.90 | 0.88 | 0.90 | 0.86 |
| NN* | 0.88 | 0.85 | 0.89 | 0.85 |

*Note\*: we use a training 70% - test 30% for NN*

In the validation set, XGB showed the strongest performance with a precision score of 0.88, F1 score of 0.86, and an accuracy score of 0.90.

The RF model showed the strongest performance on the test set with a precision score of 0.85, F1 score of 0.83, and an accuracy score of 0.88.

Neural networks perform best with large data size and as such, we restrict the data split for the neural network task to a training 70% - test 30% split. The grid search result on the NN gave best parameter values of *batch size = 8*, *epochs = 10*, and *learning rate = 0.01*.

The confusion matrices from each model gives us more information on robustness and model performances.

**Table 3**: LR Confusion Matrix on the validation set

| | | Actual | |
|---|---|---|---|
| | | Non-Hit | Hit |
| Predicted | Non-Hit | 369 | 2 |
| | Hit | 41 | 1 |

**Table 4:** DT Confusion Matrix on the validation set

| | | Actual | |
|---|---|---|---|
| | | Non-Hit | Hit |
| Predicted | Non-Hit | 330 | 41 |
| | Hit | 33 | 9 |

**Table 5**: RF Confusion Matrix on the validation set

| | | Actual | |
|---|---|---|---|
| | | Non-Hit | Hit |
| Predicted | Non-Hit | 368 | 3 |
| | Hit | 39 | 3 |

**Table 6**: XGB Confusion Matrix on the validation set

| | | Actual | |
|---|---|---|---|
| | | Non-Hit | Hit |
| Predicted | Non-Hit | 370 | 1 |
| | Hit | 40 | 2 |

**Table 7**: NN Confusion Matrix on the validation set based on

| | | Actual | |
|---|---|---|---|
| | | Non-Hit | Hit |
| Predicted | Non-Hit | 542 | 6 |
| | Hit | 65 | 6 |

*Note\*: we use a training 70% - test 30% for NN*

The DT model performed poorly with a relatively high combined false positive and false negative rate, which we consider as undesirable.

The ROC curve for the LR model is presented in figure 3.

**Figure 3:**

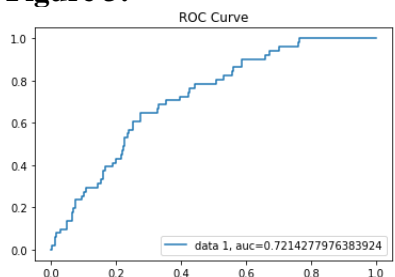

*Figure 3: ROC curve for LR model*

**Figure 4**:

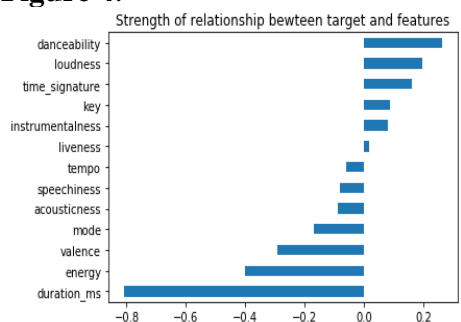

*Figure 4: Strength of relationship between target and features*

## VI. CONCLUSION AND FUTURE WORK

Using criteria for evaluating classification tasks with class imbalance, the RF, LG, and XGB, NN appears to be robust and could potentially prove useful for anyone in the music industry looking to predict Afrobeats hit songs.

We proactively decided against a three-way data split for the NN model to guide against underfitting and high bias.

Reducing cutoff point for hit songs to correct the inherent class imbalance in the dataset from 47 to 25 increased the performance of the NN model.

In future experiments, we would like to include the *section* feature provided in Spotify's *Audio Analysis* endpoint to alongside a larger dataset and a lower popularity base value to reevaluate model performances.

## REFERENCE

[1] Georgieva, E., Suta, M., & Burton, N. (2018). Hitpredict : Predicting Hit Songs Using Spotify Data Stanford Computer Science 229 : Machine Learning.

[2] Middlebrook, Kai & Sheik, Kian. (2019). Song Hit Prediction: Predicting Billboard Hits Using Spotify Data.

[3] The Secret Hit-Making Power of the Spotify Playlist https://www.wired.com/2017/05/secret-hit-making-power-spotify-playlist/

[4] Spotify Track Documentation: https://developer.spotify.com/documentation/web-api/reference/tracks/get-track/

[5] Spotify Audio-Features Documentation: https://developer.spotify.com/documentation/web-api/reference/tracks/get-audio-features/

[6] Guyon, I. (1997). A Scaling Law for the Validation-Set Training-Set Size Ratio.

[7] XGBoost: A Scalable Tree Boosting System arXiv:1603.02754 [cs.LG]

[8] Jeni, L. A., Cohn, J. F., & De La Torre, F. (2013). Facing Imbalanced Data Recommendations for the Use of Performance Metrics. International Conference on Affective Computing and Intelligent Interaction and Workshops: [proceedings]. ACII (Conference), 2013